\documentclass[aps,eqsecnum,twocolumn,showpacs]{revtex4}

\usepackage{graphicx,amsmath,amsfonts,amssymb,bm}

\begin{document}

\title{Classical momentum diffusion in double-$\delta$-kicked particles}

\author{M.M.A. Stocklin and T.S. Monteiro}

\affiliation{Department of Physics and Astronomy, University College
London, Gower Street, London WC1E 6BT, U.K.}

\date{\today}

\begin{abstract}
We investigate the classical chaotic diffusion of atoms subjected 
to {\em pairs} of closely spaced
pulses (`kicks') from standing waves of light (the $2\delta$-KP).
Recent experimental studies with cold atoms implied an underlying classical
 diffusion of a type very different from the well-known 
paradigm of Hamiltonian chaos,  the Standard Map.
 The kicks in each pair are separated by a small time
interval $\epsilon \ll 1$, which together with the kick strength $K$,
characterizes the transport. Phase space for
the $2\delta$-KP is partitioned into momentum `cells' partially
separated by momentum-trapping regions where diffusion is slow. 
We present here an analytical derivation of the classical diffusion for a
$2\delta$-KP including all important correlations which were used to
analyze the experimental data.
 We find a new
asymptotic  ($t \to \infty$) regime of `hindered' diffusion:
while for the Standard Map the diffusion rate, for $K \gg 1$, 
$D \sim K^2/2[1- 2J_2(K)...]$ oscillates about the uncorrelated
rate $D_0 =K^2/2$, we find analytically, that the $2\delta$-KP
can equal, but never diffuses faster than, a random walk rate.
 We argue this is due to the destruction of the important 
classical `accelerator modes' of the Standard Map.  
 We analyze the experimental regime $0.1\lesssim K\epsilon \lesssim 1$, where
quantum localisation lengths $L \sim \hbar^{-0.75}$ are affected by fractal
cell boundaries. We find an approximate asymptotic diffusion rate $D\propto
K^3\epsilon$, in correspondence to a $D\propto K^3$ regime in the
Standard Map associated with  `golden-ratio' cantori.
\end{abstract}


\maketitle

\section{INTRODUCTION}

The `$\delta$-kicked particle' ($\delta$-KP) is one of the best
known examples of classical Hamiltonian chaos. A particle, or in an
experiment usually a large ensemble of ultracold atoms, is
periodically `kicked' by a series of very short laser pulses forming
standing waves of light. The effective potential takes a sinusoidal
form $V(x,t)=-K\cos x \sum_N \delta (t-NT)$. Here $T$ is the kicking
period, while $K$ is the kick strength, related to the laser
intensity. The classical dynamics for the $\delta$-KP are given by
the textbook example of chaos which is the `Standard Map' \cite{Ott}.
For large $K$, the dynamics are characterized by diffusion in momentum.
To lowest order, this represents a random walk in momentum , hence
$<p^2> \sim D_0 t$, where $D_0 \sim K^2/2$.
 The quantum counterpart of the $\delta$-KP is the quantum kicked particle
 (QKP). It has also been
extensively investigated in numerous theoretical (see eg
\cite{Shep,Casati}) and experimental studies in Austin
 \cite{Moore}, Auckland \cite{Nielsen}, Oxford \cite{Burnett},
Lille \cite{Garreau}, and Otago \cite{Duffy}.

However, a recent experimental and theoretical study \cite{PRL} of
cold cesium atoms exposed to closely spaced {\em pairs} of pulses
(the 2$\delta$-kicked particle) showed chaotic classical diffusion
quite different from all other previously studied $\delta$-kicked
systems. The two kicks in each pair are separated by a short time
interval $\epsilon << T$. 

The cold cesium atoms are of course a realization of the quantum
counterpart of the $2\delta$-KP; like the single-kick QKP the
experiment exhibits the quantum chaos phenomenon of dynamical
localization \cite{Shep,Casati,Fishman} whereby the quantum
diffusion is arrested at a characteristic timescale, the
`break-time', $t^*\sim K^2/\hbar^2$. The momentum distribution of
the atomic cloud is `frozen' for times $t>t^*$ with a momentum
localization length $<p^2>\sim L^2$. By adjusting $\epsilon$ and hence the
timescales of the diffusion correlations, relative to the
break-time, it was found in \cite{PRL} that the experiment probed
distinct diffusive regimes characterized by different `families'
of long-ranged correlations. In the Standard Map, the main corrections
to the uncorrelated random walk come from 2 and 3-kick correlations.
For the $2\delta$-KP there were also families of terms correlating all kicks. 
 Hence these additional $2\delta$-KP corrections were termed `global correlations'. 

\begin{figure}[htb]
\includegraphics[width=1.5in]{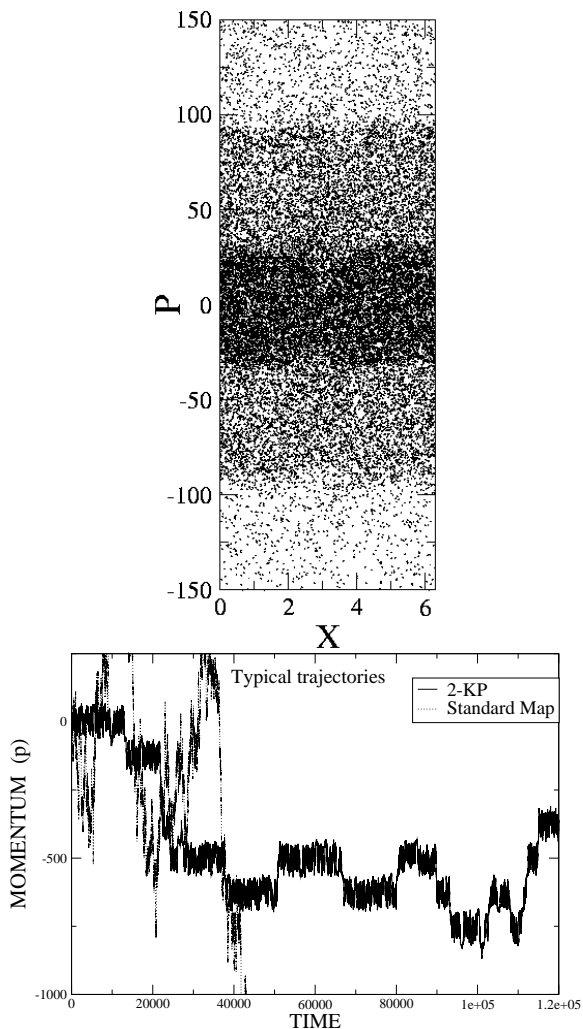}
\includegraphics[width=3.0in]{Fig1b.eps}
\caption{(top) Surface of section plot for the $2\delta$-KP, $K=7$,
$\epsilon =0.05$, $\tau=T-\epsilon =1.95$, for atoms all with
initial momentum $p_0=0$. The cellular structure is evident:
momentum space is divided into regions of fast momentum diffusion
separated by porous boundaries, ie narrow trapping regions where
classical trajectories `stick' for relatively long periods. The
trapping regions are at momenta $p \simeq \pm(2m+1)\pi/\epsilon$ where
$m=0,1,2,..$.(bottom) A typical trajectory  of the $2\delta$-KP compared
with a Standard Map trajectory: the $2\delta$-KP
trajectory spends considerable time trapped in a cell before escaping
onto the next; the Standard Map trajectory looks like a simple random walk.} \label{Fig1}
\end{figure}

 A particularly
interesting feature of the $2\delta$-KP is the `cellular' structure of
the classical phase space. This structure is analyzed in detail
below, but the basic idea is illustrated
 in Fig.\ref{Fig1}. An ensemble of particles, all
initially with momentum $p=p_0$ diffuse chaotically, but it is seen
that the diffusion is hindered at `trapping regions' ie regions with momenta
$p \simeq \pm(2m+1)\pi/\epsilon$ and $m=0,1,2,..$. For small $K\epsilon$
trajectories spend most of their time stuck in the trapping regions.
 Once the particles have escaped, the timescale
taken to diffuse freely over the remainder of the cell is
negligible.

\begin{figure}[htb]
\includegraphics[width=3.5in]{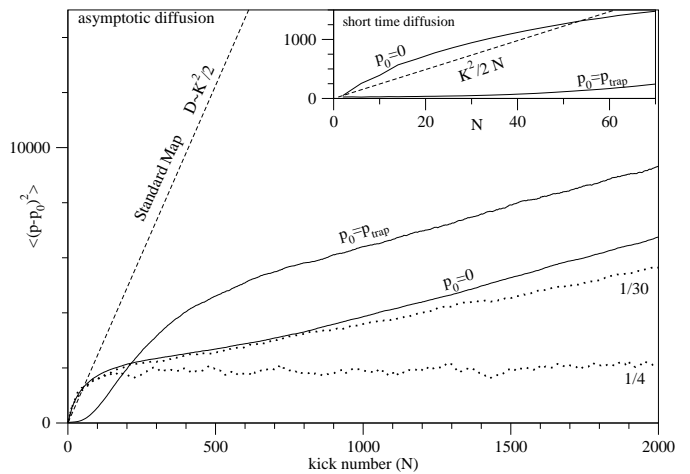}
\caption{Classical diffusion of the Standard Map vs the
$2\delta$-KP. Here $K=7$ in both cases. For the Standard Map (broken
lines) for $K \gg 1$, the same  diffusion rate $ \sim K^2/2$ characterizes  all
timescales and all starting conditions. For the $2\delta$-KP ($\epsilon=0.05$), the
diffusion rate $D(p_0,t)$ at first  depends on the initial momentum $p_0$ and time.
The inset magnifies the  $N < 50$  behavior. For atoms prepared near
the cell boundaries, ie $p_0 =p_{trap}= (2m+1)\pi/\epsilon$, where
$m=0,1,2...$, we find $D(p_0=p_{trap},t) \simeq 0$ for the first 20
or so kicks, but then diffusion speeds up rapidly. For atoms
prepared at the center of the cells, $p_0=2m\pi$, there is initially
rapid diffusion until the ensemble reaches the cell boundaries.
 We see that as $t\to \infty$, (in this case, for $N \gtrsim 200$) the
diffusion rate reaches an asymptotic value $ D(p_0,t) \equiv D_{\infty}$
independent of $p_0$ . But $D_{\infty} < D_0$:
 the diffusion rate is always slower than the uncorrelated Standard Map.
 The quantum behavior is illustrated by the dotted lines. For effective
$\hbar=1/4$, the quantum diffusion follows the classical behavior
for only a few kicks before it is arrested by Dynamical
Localization. For smaller $\hbar \simeq 1/30$ it follows it for
considerably longer.} \label{Fig2}
\end{figure}

The experiments in \cite{PRL} showed that for short and intermediate times ($\sim
100$ kicks for the experimental parameters) the diffusion rates
depended strongly on $p_0$ and $t$. Here we examine also for the
first time the asymptotic regime ($t\to\infty$) where the momentum
spread of the atomic cloud is large compared with a single cell. At
very long times momentum-dependent correlations decay to zero,
leaving linear diffusion corrections leading to $<p^2>=D_{\infty}t$.
 Fig.\ref{Fig2} provides a summary of the short-time and the asymptotic
momentum diffusion regimes in the $2\delta$-KP.

The regime $0.1 \lesssim K\epsilon \lesssim 1$ is of special
interest because it corresponds roughly to the experimental parameters
and because  a new study of the {\em quantum} $2\delta$-KP
\cite{critstats,QKR2} found a scaling behavior quite different from the usual
QKP for this regime: while for the QKP the localization length
$L\sim\hbar^{-1}$, for the $2\delta$-KP, $L\sim\hbar^{-0.75}$. To
date this result is not fully explained. Here, we find 
$D_{\infty}\sim K^3\epsilon$ in this regime.

The structure of the paper is as follows:\\
{\bf(a)} In Section II  we review the classical dynamics of the Standard
Map and the $2\delta$-KP, we explain the cellular structure
 of the $2\delta$-KP, we show it is natural to work in rescaled
 momenta $p_{\epsilon}=p \epsilon$ to remove effects dependent on
 the cell size $2\pi/\epsilon$ upon which there is an effective kick strength 
$K_{\epsilon}=K \epsilon$ and 
we obtain an approximate map for the trapping regions
 and present a heuristic justification for the
$D_{\infty}\sim K^3\epsilon$ behavior which is seen in the $K \epsilon \lesssim 1$
regime (where fractional power-law behavior was found for
the quantum localization lengths $ L \sim \hbar^{-0.75}$ \cite{critstats,QKR2}).\\
{\bf(b)} In Section III we explain how to derive
the form of the long-range correlations which modify the diffusion
rate of the $2\delta$-KP. While these results were used to analyze
the experiments in \cite{PRL}, due to space constraints a full
derivation could not be presented there. The emphasis is on timescales
comparable to the quantum break-time $t^*$: since the diffusive behavior
is `frozen-in' at this point, even transient (but long-lived)
classical correlations become important for the quantum dynamics and hence are
essential to understand
the experimental data. This section is the most technical
however, and readers without a detailed interest in classical correlations
can obtain the key points from Figs.5-7 and their captions.\\
{\bf (c)} In Section IV we investigate for the first time the asymptotic
linear diffusion regime ($t\to\infty$). We attempt to derive analytical forms for $D_{\infty}$.
For large $K \epsilon$, one of the main results of the present work
is the formula $D_{\infty} \simeq K^2 [1-\frac{J_1^2(K\epsilon)}{1-J_0^2(K\epsilon)}]$
which gives extremely accurate results compared with numerics (see Fig.\ref{Fig9}). It is shown from this that
the diffusion rate  of the $2\delta$-KP can approach, but not exceed the 
 uncorrelated rate (ie the value expected if the particles executed
a random walk). We attribute this to the absence of accelerator modes \cite{Fishaccel}
which have been observed experimentally for the usual QKP \cite{Raizen} but are not found in 
the $2\delta$-KP. For $K \epsilon \lesssim 1$, although we show that
including higher-order families of
long-ranged correlations makes the analytical diffusion {\em tend} to the numerical
result we are unable to quantitatively reproduce the $D_{\infty}\sim K^3\epsilon$ behavior.
{\bf(d)} Finally, in Section V, we conclude.

\section{CLASSICAL DYNAMICS}

\subsection{Standard Map}

The classical Standard Map is obtained by integrating Hamilton's
equations for motion in the $\delta$-kick potential $V(x,t)=-K\cos x
\sum_N \delta (t-NT)$. One obtains two equations which may be solved
iteratively to evolve the system through each period $T$:
\begin{eqnarray}
p_{N+1}=p_N + K\sin x_N; &\ & x_{N+1}=x_N + p_{N+1}T \nonumber \\
\label{eq1}
\end{eqnarray}
A $\delta$-kick is followed by a period of free evolution with
constant momentum. With increasing kick-strength $K$ the system
makes a transition from integrable, regular dynamics to eventual
full chaos. For $K\gtrsim 1$, chaotic momentum diffusion is
unbounded and all chaotic phase-space regions are connected. If we
neglect all correlations between impulses, ie assuming
$\textless\sin x_N \sin x_{N'}\textgreater\simeq0$ for all kicks,
the momentum of a trajectory in effect represents a `random walk'.
The corresponding energy of an ensemble of particles grows linearly
with time, since  $<p^2>= D_0 N\simeq K^2/2 N$. {\em NB:}
from Sec.III.B onwards, time is measured in kick-{\em pairs} $t=N/2$; in
that case all diffusion rates are doubled; eg  $<p^2> \sim D_0 t = K^2 t$;
hence the uncorrelated diffusion rate in those units would be given
by $D_0 = K^2$.

The overall diffusion in the chaotic regime is in general however
not uncorrelated (provided $K$ is not too large). In
\cite{RechWhite}, the effect of short-range correlations between
kicks was investigated theoretically. A more accurate form for the
diffusion rate $D\simeq K^2(\frac{1}{2}-J_2(K)-J_1^2(K)...)$ was
obtained, where $J_m(x)$ is a regular Bessel function of the first
kind of order $m$ and argument $x$. The second term is a 2-kick
correction resulting from correlations $<\sin x_N \sin x_{N+2}>$;
the third term is a 3-kick correction resulting from $<\sin x_N \sin
x_{N+3}>$. The effects of these corrections on the energy absorbed
by atoms in pulsed optical lattices have been experimentally
observed \cite{Raizen}. Note that for the Standard Map, the
correlations represent a simple change in the magnitude of $D$; the
energy increase is still linear in time. In \cite{PPRtheory} it was
further shown that in an asymmetric potential, the 2-kick
correlations yield a local correction to the diffusion, ie $D$
depends on both time and the relative initial momentum, $p_0$,
between the atoms and the standing wave of light. This produces a
type of chaotic Hamiltonian ratchet.

\subsection{The $2\delta$-KP}

The classical map for the $2\delta$-KP is a straightforward
extension of the Standard Map:
\begin{eqnarray}
p_{N+1}=p_N + K\sin x_N; &\ & p_{N+2}=p_{N+1} + K\sin x_{N+1}
\nonumber \\
x_{N+1}=x_N + p_{N+1}\epsilon; &\ & x_{N+2}=x_{N+1} + p_{N+2}\tau
\nonumber \\
\label{eq2}
\end{eqnarray}
where $\epsilon$ is a very short time interval between two kicks in
a pair and $\tau$ is a much longer time interval between the pairs.
It is easily seen from the map that atoms for which
$p_0\epsilon=(2m+1)\pi$ and $m=0,1,2,...$ experience an impulse
$K\sin x_0$ followed by an impulse $\simeq K\sin(x_0+\pi)$ which in
effect cancels the first. The regime $p_0\simeq(2m+1)\pi/\epsilon$
corresponds to the `momentum trapping' regions. Conversely in the
case $p_0\epsilon=2m\pi$, a series of near-identical kicks produces
initially rapid energy growth.

Some of the characteristics of the diffusion can be
analyzed  by the properties of the classical map in the
trapping regions. Starting from the map (\ref{eq2}) with $N=0$ we
re-scale all variables $p^{\epsilon} = p\epsilon$ and $K_{\epsilon}
= K \epsilon$, $\tau_{\epsilon}= \tau/\epsilon \gg 1$ to obtain
\begin{eqnarray}
p^{\epsilon}_1=p^{\epsilon}_0 + K_{\epsilon}\sin x_0; &\ &
p^{\epsilon}_2=p^{\epsilon}_1 + K_{\epsilon}\sin
x_1 \nonumber \\
x_1=x_0 + p^{\epsilon}_1; &\ & x_2=x_1 +
p^{\epsilon}_2\tau_{\epsilon} \label{eq3}
\end{eqnarray}
We take $p^{\epsilon}_0=p^{\epsilon}_R + \delta p^{\epsilon}$, where
$p^{\epsilon}_R$ is the trapping momentum, ie $p_R^{\epsilon m} =
(2m+1)\pi$ and we choose $m=0$ for the first trapping region.
Inserting the above into \ref{eq3} we have
\begin{eqnarray}
p^{\epsilon}_2=p^{\epsilon}_0+K_{\epsilon}\sin x_0+K_{\epsilon}\sin
(x_0+\pi+\delta p^{\epsilon} +
K_{\epsilon}\sin x_0) \nonumber \\
=p^{\epsilon}_0+K_{\epsilon}\sin x_0 +
K_{\epsilon}[\sin(x_0+\pi)\cos(\delta p^{\epsilon} +
K_{\epsilon}\sin x_0) \nonumber \\
+ \cos(x_0+\pi)\sin(\delta p^{\epsilon} +
K_{\epsilon}\sin x_0)] \nonumber \\
\label{eq4}
\end{eqnarray}
Assuming small angle identities throughout, $\cos(f(\epsilon))\simeq
1$ and $\sin(f(\epsilon))\simeq f(\epsilon)$,
\begin{eqnarray}
p^{\epsilon}_2\simeq p^{\epsilon}_0-K_{\epsilon}\cos x_0[K_{\epsilon}\sin x_0
+ \delta p^{\epsilon}]\nonumber \\
=p^{\epsilon}_0 - \frac{K_{\epsilon}^2}{2}\sin 2x_0 -
K_{\epsilon}\delta p^{\epsilon}\cos x_0 \label{eq5}
\end{eqnarray}
So at the center of the trapping region ($\delta p^{\epsilon} = 0$)
the double-kick map is equivalent to an effective $\sin 2x$ single
kick Standard Map:
\begin{eqnarray}
p_2^{\epsilon R}\simeq p^{\epsilon}_0 - K_R\sin 2x_0 \label{eq6}
\end{eqnarray}
where $K_R=K^2_{\epsilon}/2$. Further away from the exact trapping
momentum where $\delta p^{\epsilon} >> K_{\epsilon}/2$ we have a
cosinusoidal map:
\begin{eqnarray}
p_2^{\epsilon \delta p}\simeq p^{\epsilon}_0 - K_{\delta p} \cos x_0
\label{eq7}
\end{eqnarray}
where $K_{\delta p}=K_{\epsilon}\delta p^{\epsilon}$.

 From these arguments we clearly see that the natural stochasticity  parameter
of the $2\delta$-KP in re-scaled momenta $p^{\epsilon}=p\epsilon$ is
$K_{\epsilon}=K\epsilon$. It is also important to estimate the range of $0<|\delta
p^{\epsilon}|<\delta p^{\epsilon}_{max}$: the trapping regions have
a small but finite width determined by whether there is significant
cancelation between consecutive kicks. Requiring $K_{\epsilon}\sin
x_0 + K_{\epsilon}\sin (x_0+\pi + \delta p^{\epsilon}) \simeq 0$ we
estimate $\delta p^{\epsilon}_{max}\simeq \pi/6$, in other words over about a sixth 
of the 
width of each momentum cell, classical
trajectories experience significant trapping in the momentum
diffusion.

\begin{figure}[htb]
\includegraphics[width=3.0in]{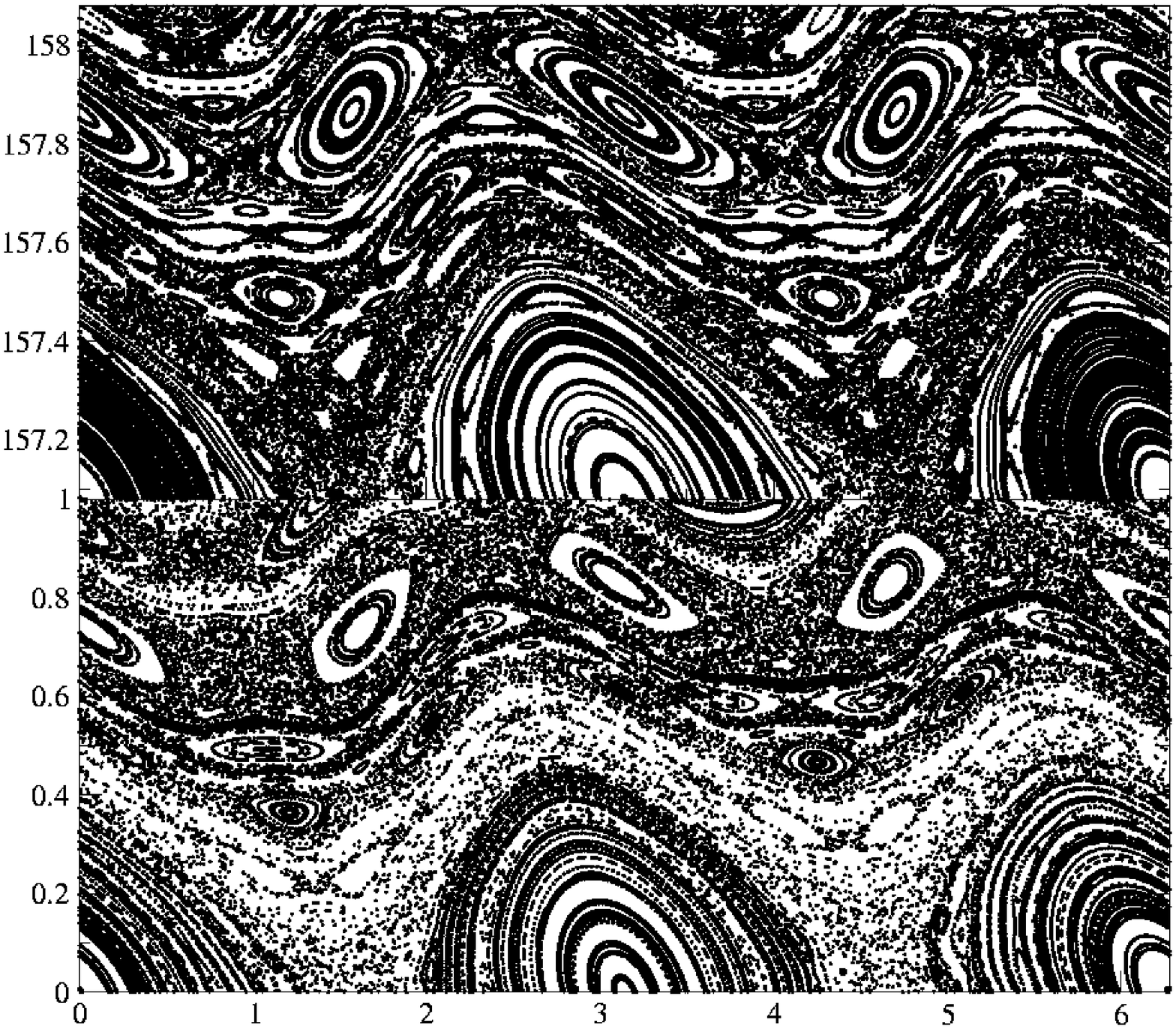}
\caption{Correspondence between the local phase space at the
resonant momentum $p_R$ of a trapping region ($K=4.8$, $\epsilon
=0.02$, $\tau=1.98$) in the double-kicked system (top) and a
standard map (bottom) with a kick $p_2=p_0 - K_R\sin 2x_0$, where
$K_R=\frac{K^2\epsilon}{2}=0.24$ is the kick strength of the
Standard Map and the period is $T=2$. Note that unscaled momenta are
used here.} \label{Fig3}
\end{figure}

\begin{figure}[htb]
\includegraphics[width=3.0in]{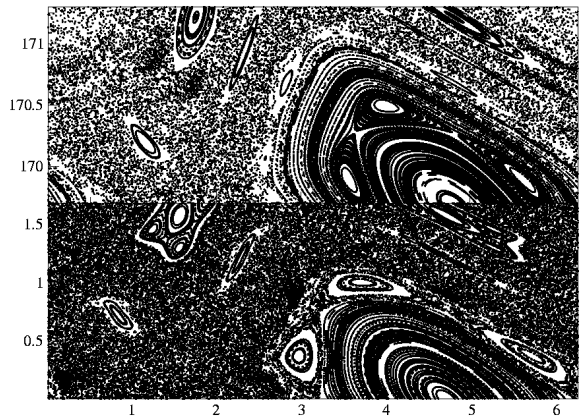}
\caption{Correspondence between the outer phase space near the
resonant momentum $p_R$ of a trapping region ($K=4.0$, $\epsilon
=0.02$, $\tau=1.98$) in a double-kicked system (top) and a $\pi/2$
phase-shifted cosinusoidal standard map (bottom) $p_2=p_0 -
K_{\delta p}\cos x_0$, where $K_{\delta p}=K\epsilon\delta p=0.75$
is the kick strength of the Standard Map and the period is $T=2$.
Note that unscaled momenta are used here.} \label{Fig4}
\end{figure}

In Figures \ref{Fig3} and \ref{Fig4} we can see the change from a
$\sin 2x$ map at $p^{\epsilon}\simeq p^{\epsilon}_R$ where we
observe two sets of stable islands within the range $0<x<2\pi$, to a
cosinusoidal map further out where the position of islands is
shifted by a phase of $\pi/2$. Fig.\ref{Fig3} compares the detailed
structure at $p^{\epsilon}\simeq p^{\epsilon}_R$ with a standard map
phase space for which $K_{SM}=K^2\epsilon/2$. Fig.\ref{Fig4}
compares an enlarged section of a `side resonance' further away from
$p^{\epsilon}_R$, with a standard map with $K_{SM}=K\epsilon\delta
p$. In both cases the correspondence is clearly visible.

We note that in the Standard Map there is a transition to unbounded
diffusion for a critical kick strength $K=K_{crit}\simeq 1$ where
the last invariant curve breaks, leaving a fractal cantorus
structure. This curve corresponds to momenta $p\simeq 2\pi m
\mathbb{R}$, where $\mathbb{R}$ is the golden ratio. In
\cite{scaling} it was shown that for $K_{crit}< K \lesssim 4.5$ the
diffusion rate is $D\simeq 0.3(K-K_{crit})^{\eta}$. The value
$\eta\simeq 3$ was obtained in \cite{scaling} from the scaling
properties of the classical map near the golden-ratio cantorus.

From the above, one can hazard a rough order-of-magnitude
justification for a cubic dependence of the asymptotic diffusion rate $D_{\infty}
\simeq K^3\epsilon$ for $0.1  \lesssim K_{\epsilon} \lesssim 1$ 
found numerically (see Sec.IV). Note that in scaled momenta,
$D_{\infty} \sim K_{\epsilon}^3$.
  We note the correspondence
between the Standard Map and the $2\delta$-KP with $K\to
K_{\epsilon}$; for small $K_{\epsilon}$, $\delta p^{\epsilon}
> 0$, the form of the map in \ref{eq7} is dominant, since
trajectories spend most of their time in the trapping regions. In
this case, the scaled critical kick strength $K_{\epsilon}^{crit}
\simeq \epsilon$. Taking a representative value of $<\delta
p^{\epsilon}> \simeq \delta p^{\epsilon}_{max}/2 =\pi/12 \simeq
1/4$, the effective kick strength in \ref{eq7} is $\simeq
K_{\epsilon}/4$. We thus have a diffusion rate {\em in the trapping
regions} in terms of re-scaled momenta $<p^2_{\epsilon}> \sim 0.3
(\frac{K_{\epsilon}}{4}-K_{\epsilon}^{crit})^3 t$. Then, in unscaled
momenta, we would have $D\sim K^3\epsilon$ provided $K$ is reasonably
large compared with $\epsilon$. While this is by no means a rigorous
argument it may provide some indication of the mechanism
underpinning the $K^3$ dependence. We note  the $0.75$ exponent in
the $L \propto \hbar^{-0.75}$ scaling is also an exponent associated
with scaling properties of the region around the Golden ratio
cantori. We propose tentatively this supports the proposal in
\cite{QKR2,critstats} that the properties of the  $2\delta$-KP are intimately
connected with the properties of the Golden ratio cantorus.

\section{MOMENTUM DIFFUSION}

\subsection{Standard Map}

The classical diffusion corrections for the Standard Map were first
obtained by Rechester and White \cite{RechWhite} and we follow their
notation closely. From the map (\ref{eq1}), the momenta and
positions of a trajectory evolve by a sequence of impulses: $p_N =
p_0 + S_{N-1}$ and $x_N = x_{N-1}+p_0+S_{N-1}$, where
$S_l=\sum^l_{j=0} K\sin x_j$, the initial momentum of an atom is
$p_0$ and the period $T$ is taken to be unity. If we consider an
ensemble of particles with an initial probability distribution in
position and momentum $G(x_0,p_0,t=0)$, at a later time (measured in
number of kicks $t=N$) the distribution is given by
\begin{widetext}
\begin{eqnarray}
G(x_t,p_t,t)=\sum^{+\infty}_{n_t=-\infty}...\sum^{+\infty}_{n_1=-\infty}
\int^{2\pi}_0 dx_0dp_0\ G(x_0,p_0,0) \int^{2\pi}_0 dx_t \ ...
\int^{2\pi}_0 dx_1\ \delta(p_t-p_0-S_{t-1}) \nonumber \\
\times \ \delta(x_t-x_{t-1}-p_0-S_{t-1}+2\pi n_t) \ ...\
\delta(x_1-x_0-p_0-S_0+2\pi n_1) \label{eq8}
\end{eqnarray}
\end{widetext}
The sums over $n_1\ ...\ n_t$ appear because of the periodicity of
phase space in $x_0\ ...\ x_t$. The momentum diffusion rate $D$ is
given by
\begin{eqnarray}
D=\lim_{t\to\infty}\frac{1}{t}<(p_t-p_0)^2>_t \nonumber \\
= \frac{1}{t}\int^{2\pi}_0dx_t \int^{+\infty}_{-\infty}dp_t\
G(x_t,p_t,t)(p_t-p_0)^2 \label{eq9}
\end{eqnarray}
By taking the initial distribution as $G(x_0,p_0,0)=\frac{1}{2\pi}
\delta(p-p_0)$ (ie a uniform spatial distribution with all particles
at initial non-zero momentum $p_0$) and using the Poisson sum
formula giving the Fourier transform of a $\delta$-spectrum, $\sum_n
\delta(y+2\pi n) = \frac{1}{2\pi}\sum_m e^{imy}$, we can rewrite
\ref{eq9} as
\begin{widetext}
\begin{eqnarray}
D=\lim_{t\to\infty}\frac{1}{t}\sum^\infty_{m_t=-\infty} ...
\sum^\infty_{m_1=-\infty}
\prod_{i=0}^t\int^{2\pi}_0\frac{dx_i}{2\pi}\ (S_{t-1})^2\
\exp\{\sum_{j=1}^t im_j(x_j-x_{j-1}-p_0-S_{j-1})\} \label{eq10}
\end{eqnarray}
\end{widetext}
To lowest order one can set all $m_j$ coefficients to zero, thus
eliminating all exponentials. By using the previous form of $S$ and
integrating over the sine products it is easily seen that the random
walk $D_0=K^2/2$ is recovered. Higher-order corrections to the
diffusion rate are obtained by setting certain $m_j$ coefficients to
a non-zero value; for the most dominant corrections $|m_j|=1,2$. The
integrals are solved using the relation $\exp\{\pm iK\sin
x\}=\sum^{+\infty}_{n=-\infty} J_n(K)\exp\{\pm inx\}$ and for
corrections to be non-zero, all arguments of exponentials must
vanish for $2\pi$-periodic integration. This requires pairing
exponentials with the relevant Bessel functions and sine products
included in $S$.

The main corrections to the Standard Map are the 2-kick and 3-kick
correlations found in \cite{RechWhite} and account in large measure
for the experimental oscillations seen in \cite{Raizen}. The 2-kick
correlation is obtained from setting $m_j=\pm 1$ and $m_{j-1}=\mp 1$
and the lowest order 3-kick correlation from $m_j=\pm 1$ and
$m_{j-2}=\mp 1$. From the $\sin x_j \sin x_{j-2}$ and $\sin x_j \sin
x_{j-3}$ terms we obtain $C_2=-K^2J_2(K)$ and $C_3=-K^2J^2_1(K)$
respectively, as previously. There is also a higher-order 3-kick
correlation, $C_3^{'}=+K^2J^2_3(K)$, found in \cite{RechWhite} derived
from $m_j=\pm 1$, $m_{j-1}=\mp 2$, $m_{j-2}=\pm 1$, and a 4-kick
correlation, $C_4=+K^2J^2_2(K)$, cited in \cite{LichLieb} derived from
$m_j=\pm 1$, $m_{j-1}=m_{j-2}=\mp 1$, $m_{j-3}=\pm 1$. This leads to a
total correction to the diffusion of
\begin{eqnarray}
D=K^2[\frac{1}{2}-J_2(K)-J^2_1(K)+J^2_3(K)+J^2_2(K)]
\label{eq11}
\end{eqnarray}
These terms represent the correlations between two given impulses
only, two, three or four kicks apart. In the next section we shall
see that in the double-kicked system there are entire families of
terms representing correlations between a given impulse $\sin x_j$
and every other impulse. Such a global correlation family does in
fact exist for the Standard Map, as the above method of derivation
can be extended to any order of k-kick correlations between any two
given kicks \cite{map} ($\sin x_j\sin x_{j-k}$). For the lowest
order we have $m_j=\pm 1$ and $m_{j-k+1}=\mp 1$ and hence
$C_k=-K^2J^2_1(K)J_0^{k-3}(K)$ ($k\geq 3$). Clearly corrections
become smaller with increasing $k$. The 2-kick correlation is a
special case of this global collection of terms. Note the linear
time dependence of all individual k-kick correlations (ie
$<p^2>\propto C_kt$). The total correction to the momentum diffusion
rate due to all the above k-kick terms is $\sum_{k=3}^{\infty} C_k =
-K^2(J_1^2(K))/(1-J_0(K))$. Similar higher-order global corrections
can be found in the Standard Map of the general form $\pm K^2(\prod
J^m_p(K))/(\prod (1-J_0(K))^n)$ for some $m,n,p$. However in
practice, higher-order corrections beyond the basic terms in
\cite{LichLieb} can generally be neglected for all $K\gtrsim 5$
since they do not alter the diffusion rates significantly. For
smaller $K$ there can be significant differences;
however phase space becomes increasingly regular as $K$ decreases
and for $K \lesssim 2$ a diffusive approach is not justified. 
 In the double-kick system the
 higher order (long-range) correlations originate only in the 
thin trapping layers so a diffusive analysis is effective even in
a regime where long-range correlations are important.

It should be noted that none of these corrections depend on the
momenta of the atoms in the ensemble; they simply alter the
magnitude of the overall linear rate of energy absorption for a
given $K$. In any $\delta$-kicked system where the kicking periods
are all equal and there are no other asymmetries, effects associated
with momentum-dependent diffusion corrections are negligible. This
is due to such corrections including oscillations $\cos T p_0$ on a
comparable scale to the natural width of the initial momentum
distribution $\Delta p_0 \sim 2\pi$ of the atomic ensemble. Hence
these corrections average to zero. The 1-kick correlation $\sin x_j
\sin x_{j-1}$ between consecutive kicks for which a single $m_j=\pm
1$ only, is such a correction and is hence absent in the Standard
Map, but will be seen to contribute significantly to the
double-kicked system, where a much shorter kicking period is
introduced.

\subsection{The $2\delta$-KP}

For the $2\delta$-KP the notation of the Standard Map diffusion
equation can be changed slightly to include two closely-spaced kicks
for each time step denoted by $(1)$ and $(2)$; thus the evolution of
momentum is now in terms of pairs of kicks. Throughout this section
we work in unscaled momenta $<p^2> = Dt$ (in re-scaled momenta
$<p_{\epsilon}^2> = D\epsilon^2t$). $S$ takes on the form
$S_t^{(2)}=\sum^t_{m=1}\sum^2_{r=1}K\sin x_m^{(r)}$ \cite{map} and
we indicate explicitly the two time intervals $\tau$ and $\epsilon$,
defined as previously. Again, $D_0$ is obtained by setting all $m_j$
coefficients to zero, but it should be noted that as there are now
$2t$ variables, the form of $D_0$ changes to $K^2$ and hence
$<(p-p_0)^2>=D_0t=K^2t$ (or in scaled momenta $K_{\epsilon}^2t$).
Obviously this does not change the underlying physics; the new
formula is only due to a redefinition of time in terms of number of
pairs of kicks, $t=N/2$, for even $N$. Physical time is $t(\tau +
\epsilon)$.

\begin{widetext}
\begin{eqnarray}
D=\lim_{t\to\infty}\frac{1}{t}\sum^\infty_{m_t^{(2)}=-\infty}\sum^\infty_{m_t^{(1)}=-\infty}...
\sum^\infty_{m_1^{(2)}=-\infty}\sum^\infty_{m_1^{(1)}=-\infty}
\prod_{i=1}^t \int^{2\pi}_0\frac{dx_i^{(2)}}{2\pi}\int^{2\pi}_0\frac{dx_i^{(1)}}{2\pi}
(S^{(2)}_t)^2 \nonumber \\
\times\ \exp\{\sum_{j=1}^t [im_j^{(2)}(x_j^{(2)}-x_j^{(1)}-
\epsilon(p_0+S_j^{(1)})) + im_j^{(1)}(x_j^{(1)}-x_{j-1}^{(2)}-
\tau(p_0+S_{j-1}^{(2)}))]\} \label{eq12}
\end{eqnarray}
\end{widetext}

\subsection{Momentum-dependent Diffusion of the $2\delta$-KP }

The introduction of the short timescale $\epsilon$ results in a
significant 1-kick contribution to the diffusion, unlike in any
previously studied $\delta$-kicked system. For $m_j^{(2)}=\pm 1$ the
correlation involves $\cos p_0 \epsilon$ and Bessel functions of
argument $K\epsilon=K_{\epsilon}$, while for $m_j^{(1)}=\pm 1$ the
correlation involves $\cos p_0 \tau$ and Bessel functions of
argument $K\tau$. The latter case gives negligible contributions as
$\tau$ is the large time interval between pairs of kicks, resulting
in fast oscillations with $p_0$. We can effectively set all
$m_j^{(1)}=0$ for all momentum-dependent correlations - a valid
approximation provided $\tau >> \epsilon$.  The $m_j^{(2)}$
coefficients will be simply referred to as $m_j$ for the remainder
of the paper.

For the case of $\cos p_0 \epsilon$, the $\sin x_j^{(2)} \sin
x_j^{(1)}$ term results in a kick-to-kick correction
\begin{eqnarray}
C_1t=K^2\cos p_0 \epsilon(J_0(K_{\epsilon})-J_2(K_{\epsilon})) \nonumber \\
\times\ \sum^t_{j=1}J_0^{2j-2}(K_{\epsilon}) \label{eq13}
\end{eqnarray}
where $t=N/2$ ($N$ even) ie time is measured in pairs of kicks as
for all correlations that follow. Kicks at longer times have weaker
correlations than those at short times, however since
$J_0(K_{\epsilon})\to 1$ as $\epsilon \to 0$, the time-dependent
summation decays slowly with time and the lifetime of the overall
kick-to-kick correlation may far exceed the active running time of
an experiment. Note however that the correlation does decay to zero
eventually (as for all other momentum-dependent correlations), ie
$D(p_0,t\to\infty)=0$. Hence we do not take the infinite time limit
in \ref{eq12}; instead we calculate the quantity $Ct$. The geometric
sum in \ref{eq13} saturates to $1/(1-J_0^2)$ after a time $\sim
10/(K_{\epsilon})^2$ and hence
\begin{eqnarray}
\lim_{t\to\infty} C_1t=K^2\cos p_0 \epsilon
\frac{J_0(K_{\epsilon})-J_2(K_{\epsilon})}{1-J_0^2(K_{\epsilon})}
\label{eq14}
\end{eqnarray}
For short times $C_1t$ grows linearly with time, since for small
$K_{\epsilon}$ we have $J_0(K_{\epsilon})\simeq 1 >> J_2(K_{\epsilon})$. We
can approximate the correlation to $K^2t\cos p_0 \epsilon$ and hence
the average energy of the double-kicked particle also grows linearly
in this regime.

\begin{figure}[htb]
\includegraphics[height=3.0in,width=3.0in]{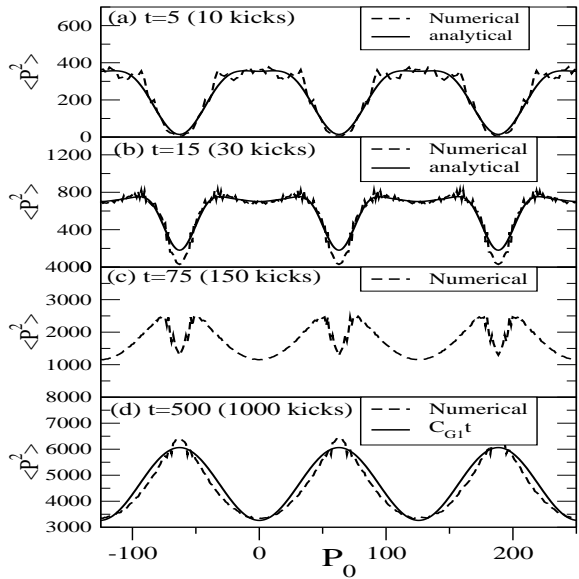}
\caption{Agreement between analytical diffusion corrections and
numerical double-kick simulation of 100,000 classical particles at
$K=7$, $\epsilon=0.05$, $\tau=1.95$, for various times measured in
pairs of kicks, $t=N/2$ where $N$ is the (even) number of individual
kicks. Physical time is $t(\tau + \epsilon)$ where $\epsilon$ is the
short time interval between kicks in a pair and $\tau$ is the long
time interval between pairs. Energy absorption is plotted as a
function of initial momentum of the particles. The corrections
included in the analytical curve in (a) are $C_1t$, $C_{G1}t$ and
the lowest order $C_{G2}^{(1)}t$ term labeled IIa in Table 1. In (b)
all correlations given in Table 1 are included. Agreement is
excellent at short times, but higher-order terms are needed at later
times. One observes a sign reversal of the cosine envelope and
gradual disappearance of the inverted Poisson peaks at initial
minima. Particles initially prepared in momentum-trapping regions
eventually absorb the most energy. At very long times the lowest
order global correlation $C_{G1}$ dominates almost completely.}
\label{Fig5}
\end{figure}

Fig.\ref{Fig5} is the same as in \cite{PRL} showing a numerical
simulation of the energy absorption of an ensemble of 100,000 classical
particles at $K=7$, $\epsilon=0.05$, $\tau=T-\epsilon=1.95$, as a
function of their initial momenta $p_0$ at various times (measured
in pairs of kicks). The numerics are superposed with combinations of
the time-dependent correlations $C_jt$ presented in this paper. In
Fig.\ref{Fig5}a the basic cosine behavior of $C_1$ is clearly visible:
atoms with initial momenta $p_0\simeq2m\pi/\epsilon=0,125.66,...$
experience the largest energy absorption, while those prepared at
$p_0\simeq(2m+1)\pi/\epsilon=62.83,188.50,...$ are `trapped' near
this initial momentum and experience almost no energy absorption.
When one looks at Figures \ref{Fig5}c and \ref{Fig5}d, something
unexpected occurs. The maxima of Fig.\ref{Fig5}a slowly turn into near
minima, while energy absorption for atoms near the `trapping' regions
increases continuously. A complete reversal of the initial situation
eventually occurs at longer times, $<(p-p_0)^2>\propto -\cos p_0
\epsilon$.

\begin{figure}[htb]
\includegraphics[width=3.5in]{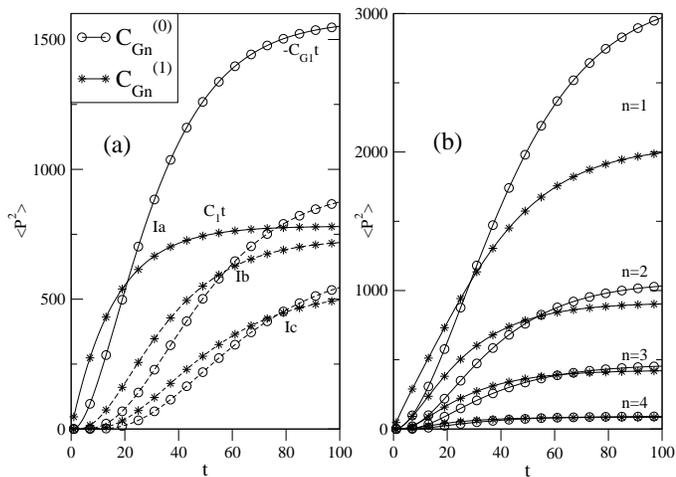}
\caption{Behavior of global correlations $C_{Gn}t$ with number of
pairs of kicks. Absolute values $|C|t$ of maximum energy absorption
are shown ($\cos np_0 \epsilon=1$). Fig.\ref{Fig6}a shows the
comparison between individual correlations at order $n=1$, including
nearest neighbor $C_1$, lowest order global $C_{G1}$ and higher
order Poisson correlations. Fig.\ref{Fig6}b shows the comparison
between different cosine orders $n$. Trapping terms $C_{Gn}^{(1)}$
are shown with stars, absorption enhancing terms $C_{Gn}^{(0)}$ with
circles. The labels correspond directly to those in Table 1 showing
explicit values at $t=15$. Terms are denoted by the cosine order $n$
resulting from the total sum of non-zero $m_j$ coefficients, and the
order in $J_1$ resulting from the number of non-zero coefficients.
Curves in \ref{Fig6}b are made up of the various correlations in
Table 1 for a given order $n$. Note the linear and quadratic rise
respectively for $C_1$ and $C_{G1}$ at early times. Absorption
enhancing terms always overtake their momentum trapping partners at
equal order after sufficient time, but the crossing point shifts to
later times as both the order $n$ and $O(J_1)$ increases. Overall
importance of terms also decreases for higher orders.} \label{Fig6}
\end{figure}

The reasons for this lie in considering a whole
family of global correlations similar to the case of the Standard
Map. The lowest order global correction originates from $m_j=\pm 1$, just
as for $C_1$, but this time we look at $\sin x_j^{(2)} \sin
x_i^{(r)}$ terms, where $r=1,2$ and $i<j$ but otherwise arbitrary.
In this way we include all k-kick correlations for $k\geq 2$, ie
correlations between the second impulse in a given pair, $\sin
x_j^{(2)}$, and impulses in all other pairs.
\begin{eqnarray}
C_{G1}t=\sum_{k=2}^{2t-1}C_kt=-2K^2\cos p_0 \epsilon J_1^2(K_{\epsilon})
\nonumber \\
\times \sum^t_{j=1} (2j-2)J_0^{2j-3}(K_{\epsilon}) \label{eq15}
\end{eqnarray}
The corrected energy is now $<(p-p_0)^2>\simeq K^2t+C_1t+C_{G1}t$.
Every individual correlation between two kicks in $C_{G1}$ is small
compared to the nearest neighbor correlation $C_1$, as
$J_1^2(K_{\epsilon}) << J_0 (K_{\epsilon})$. Importantly however, at
a given time the correlations do not get weaker with increasing $k$
but are equal in size for any non-nearest neighbor correlation. All
correlations however do become weaker with increasing time and
eventually saturate. Summing over $t$ reference impulses and $2j-2$
paired  correlations for each, results in a total correction to the
diffusion which after a given time becomes dominant relative to
$C_1$. Notice the difference in sign between the 1-kick correlation
and all others. The summation in \ref{eq15} can be seen to be the
derivative of that in \ref{eq13} and hence
\begin{eqnarray}
C_{G1}t=-4K^2\cos p_0 \epsilon J_1^2(K_{\epsilon})\nonumber \\
\nonumber \\
\times \frac{J_0(K_{\epsilon})-tJ_0^{2t-1}(K_{\epsilon})
+(t-1)J_0^{2t+1}(K_{\epsilon})}{(1-J_0^2(K_{\epsilon}))^2}\label{eq16}
\end{eqnarray}
For small $K_{\epsilon}$ it can be shown by assuming $J_0(K_{\epsilon})
\simeq 1-K_{\epsilon}$ and expanding the above to second
order binomial that $C_{G1}t$ initially increases quadratically at short
times:
\begin{widetext}
\begin{eqnarray}
1-K_{\epsilon}-t(1-K_{\epsilon})^{2t-1}+(t-1)(1-K_{\epsilon})^{2t+1}
\simeq
1-K_{\epsilon}-t[1-(2t-1)K_{\epsilon}+(2t^2-3t+1)K^2_{\epsilon}]
\nonumber \\
+(t-1)[1-(2t+1)K_{\epsilon}+(2t^2+t)K^2_{\epsilon}] =
2K^2_{\epsilon}(t^2-t) \label{eq17}
\end{eqnarray}
\end{widetext}
For the parameters in Fig.\ref{Fig5}, $C_{G1}t$ saturates to a value
approximately twice that of $C_1t$ as $t\to\infty$.
\begin{eqnarray}
\lim_{t\to\infty} C_{G1}t = -4K^2\cos p_0 \epsilon
\frac{J_0(K_{\epsilon})J^2_1(K_{\epsilon})}{(1-J_0^2(K_{\epsilon}))^2}
\label{eq18}
\end{eqnarray}
The behavior of the 1-kick and all global correlations with time is
illustrated in Fig.\ref{Fig6} (including higher-order Poisson
correlations presented later). The absolute value ($|C|t$) of the
maximum energy absorption (where $\cos p_0 \epsilon=1$) with
increasing number of kick pairs is shown. Note that all
momentum-dependent correlations have a non-linear time dependence
and saturate after sufficient time. This means that the
momentum-dependent diffusion rate approaches zero at long times,
where all particles absorb the same amount of energy irrespective of
$p_0$. At short times the effect of $C_{G1}$ is negligible compared
to $C_1$, but at later times the global correlations dominate the
diffusive process, explaining the sign reversal in Fig.\ref{Fig5}d.
The importance of the global correlations persists for values of
$K_{\epsilon}$ for which the corresponding phase space is completely
chaotic, a phenomenon not previously observed.

Fig.\ref{Fig5}b corresponds to a regime where $C_1$ and $C_{G1}$ are
of similar importance (near the crossing point in Fig.\ref{Fig6}).
Here and in Fig.\ref{Fig5}c another feature mentioned earlier
becomes evident. The initial troughs of the cosine in
Fig.\ref{Fig5}a turn into narrow downward peaks, superimposed onto
the cosine envelope. The origins of these peaks are global
correlation terms of higher cosine orders $\cos np_0 \epsilon$,
which yield near $\delta$-functions at the relevant momenta through
$\sum_n(-1)^n\cos np_0 \epsilon = \sum_m \delta(p_0 \epsilon -
(2m+1)\pi)$. Such Poisson terms arise when more than one $m_j$
coefficient in \ref{eq12} is set to $\pm 1$, and $\sum_j m_j = n$, ie
the sum of the coefficients defines the cosine order $n$. Terms with
$|m_j|>1$ include Bessel functions of increasingly higher orders;
since for $\epsilon\to 0$, $J_n(K_{\epsilon})\to 0$ more rapidly for
increasing $n>0$, such terms can be neglected here.

Derivations of higher-order global correlations are tedious, yet
reasonably straightforward. There are two distinct families, one
following the pattern of $C_1$ and the other following $C_{G1}$,
each causing different diffusive behavior. In both cases the terms
of interest involve $\sin x_j^{(2)}\sin x^{(r)}_k$ where $r=1,2$ and
$k<j$. In the former case $m^{(2)}_k=\pm 1$ and correlations involve
a $J_0-J_2$ factor as for $C_1$. In the latter case $m^{(2)}_k=0$
and correlations involve a $2J_1$ factor as for $C_{G1}$. The
behavior depends on whether the correlation is with a kick in a pair
associated with a corresponding zero or non-zero $m_k$ coefficient.
We choose to denote the global Poisson terms as $C_{Gn}^{(0)}$ and
$C_{Gn}^{(1)}$ depending on the family they belong to, where $n$ is
the cosine order. $C_{G1}$ is the lowest order global correlation
family belonging to $C_{G1}^{(0)}$. All Poisson terms have the
following general forms for some $m$

\begin{widetext}
\begin{eqnarray}
C^{(1)}_{Gn}t=(-1)^{n-1}K^2\cos np_0 \epsilon J_1(n K_{\epsilon})
[\prod_m J^2_1(m K_{\epsilon})][\sum_m\frac{J_0(m
K_{\epsilon})-J_2(m K_{\epsilon})}{J_1(m
K_{\epsilon})}]\sum^t_j\sum_{\alpha_m=0}^{f(j)}
(\prod_m J_0^{2\alpha_m}(m K_{\epsilon})) \label{eq19} \\
C^{(0)}_{Gn}t=(-1)^n2K^2\cos np_0 \epsilon J_1(n K_{\epsilon})
[\prod_m J^2_1(m K_{\epsilon})]
\sum^t_j\sum_{\alpha_m=0}^{f(j)}(\sum_m 2\alpha_m\frac{J_1(m
K_{\epsilon})}{J_0(m K_{\epsilon})})(\prod_m J_0^{2\alpha_m}(m
K_{\epsilon})) \label{eq20}
\end{eqnarray}
\end{widetext}
where $f(j)$ depends on the particular term:
some examples of Poisson terms are given in the appendix. Each
additional $m_k=\pm1$ coefficient adds two multiplicative $J_1$
factors of possibly different arguments to both Poisson families,
since Bessel function arguments can now be any integer multiple of
$K_{\epsilon}$, depending on the exact combination of $m_k$
coefficients. This continuously decreases the significance of the
correlations, since $J_1(x)\to 0$ as $x\to 0$. Hence the dominant
corrections are those of lowest order in $J_1$ (and with few
non-zero $m_k$ coefficients). Note that for Poisson terms other than
$C_{G1}$ various k-kick correlations are of similar but not
necessarily equal strength at a given time $t$.

For $n=1$ there are higher-order Poisson terms (in $J_1$)
contributing only to the cosine envelope in Fig.\ref{Fig5}. These
are shown in Fig.\ref{Fig6}a in comparison to the lowest order
corrections $C_1t$ and $C_{G1}t$. Again absolute values of maximum
energy absorption are shown. As $O(J_1)$ increases, terms become
less significant. For all terms, $C^{(0)}_{G1}$ correlations
eventually become more dominant than $C^{(1)}_{G1}$ correlations,
but the crossing point shifts to later times with increasing
$O(J_1)$. The difference in saturation energies between the two
types of correlations also becomes smaller.

Fig.\ref{Fig6}b shows the comparison between the combined $n=1$
correlation families and higher cosine orders. The behavior with
increasing $n$ is as before; $C_{Gn}^{(0)}$ correlations always
become more dominant than their partners of equal order, after
sufficient time. $C_{Gn}^{(1)}$ terms increase as $\sim t^n$, while
$C_{Gn}^{(0)}$ terms increase as $\sim t^{n+1}$.

Table 1 shows the maximum energy absorption values of the dominant
Poisson terms at $t=15$ (30 kicks) corresponding to Fig.\ref{Fig5}b. The
most significant diffusion corrections are those for small $n$ and
$O(J_1)$ and importance of terms rapidly decreases with higher orders.
Terms of $O(J_1^{10})$ are about 50 times smaller than the leading
corrections.

\begin{table}
\begin{tabular}{c|c|c|c}
Term & $m_j$ pattern & $O(J_1)$ & Value ($t=15$) \\
\hline Ia ($C_1$,$C_{G1}$) & $\pm 1$ & 0,2 & +472,-355 \\
\hline Ib ($C_{G1}^{(1)}$,$C_{G1}^{(0)}$) & $\pm 1$,$\pm 1$,$\mp 1$ & 4,6
& +100,-33
\\ \hline Ic ($C_{G1}^{(1)}$,$C_{G1}^{(0)}$) & $\pm 1,\pm 1,\pm 1,\mp 1,\mp 1$ & 8,10 &
+14,-4 \\
\hline IIa ($C_{G2}^{(1)}$,$C_{G2}^{(0)}$) & $\pm 1,\pm 1$ & 2,4 & -227,+113 \\
\hline IIb ($C_{G2}^{(1)}$,$C_{G2}^{(0)}$) & $\pm 1,\pm 1,\pm 1,\mp 1$
& 6,8 & -59,+24
\\ \hline IIIa ($C_{G3}^{(1)}$,$C_{G3}^{(0)}$) & $\pm 1,\pm 1,\pm 1$ & 4,6 & +82,-39
\\ \hline IIIb ($C_{G3}^{(1)}$,$C_{G3}^{(0)}$) & $\pm 1,\pm 1,\pm 1,\pm 1,\mp 1$ & 8,10 &
+29,-12 \\
\hline IVa ($C_{G4}^{(1)}$,$C_{G4}^{(0)}$) & $\pm 1,\pm 1,\pm 1,\pm 1$
& 6,8 & -29,+14
\\
\end{tabular}
\caption{Diffusion correlations shown in Fig.\ref{Fig6}.}
\end{table}

The momentum diffusion corrections derived here now enable us to
explain the behavior seen in Fig.\ref{Fig5}, including the inverted
peaks. The agreement between the numerical data and the analytical
diffusion calculations is generally good. At very short times
(Fig.\ref{Fig5}a) essentially only $C_1$, $C_{G1}$ and
$C^{(1)}_{G2}$ (IIa) contribute to the diffusion and agreement is
excellent. In Fig.\ref{Fig5}b all of the correlations in Table 1
have been included in the analytical curve and good agreement is
obtained: the sign of the cosine envelope starts to change and the
inverted peaks slowly vanish at later times. From \ref{eq19} and
\ref{eq20} one notes that at the same order $n$, the two types of
correlations are of different sign and so oppose each other, clearly
seen in the case of $n=1$. For $n>1$ it is found that while
$C_{Gn}^{(1)}$ correlations contribute to increasing the size of the
downward peaks in all cases and thus favor momentum trapping,
$C_{Gn}^{(0)}$ correlations result in enhanced energy absorption for
initially trapped atoms. These latter correlations cause the release
of atoms from trapping regions in momentum space.

In Fig.\ref{Fig5}c the analytical curve has been omitted as
reasonable agreement cannot be achieved using only the correlations
in Table 1. Higher-order terms are needed at intermediate times,
however at very long times the diffusion is dominated by $C_{G1}$,
the most important correlation, as can also be seen in
Fig.\ref{Fig6}a. After saturation, the higher-order $n>1$ Poisson
correlations, result in superimposed upward peaks at the maxima of
the cosine envelope giving the latter a pointed appearance. The
newly found global correlation families lead to a situation where at
long times, when diffusion no longer depends on initial momentum,
those atoms that started in a trapping region have actually gained
more energy than those that started in an enhanced absorption region
in momentum space.

So far this paper has only treated the diffusion problem
classically, but a real experiment would obviously be carried out in
the quantum regime. In \cite{PRL}, \cite{Floquet} and \cite{critstats} it was shown
that the effects presented here can readily be observed in a quantum
experiment, and indeed a range of other interesting features were
discovered.

As mentioned in the introduction to this paper, energy absorption
does not continue indefinitely in the quantum case. The energy
saturates to a near-constant value after a characteristic quantum
break time $t^*\sim D/\hbar^2$ \cite{Shep}, where $\hbar$ is a
scaled Planck constant. The smaller $\hbar$, the longer the system
follows the classical predictions (see Fig.\ref{Fig2}).

\begin{figure}[htb]
\includegraphics[width=3.0in]{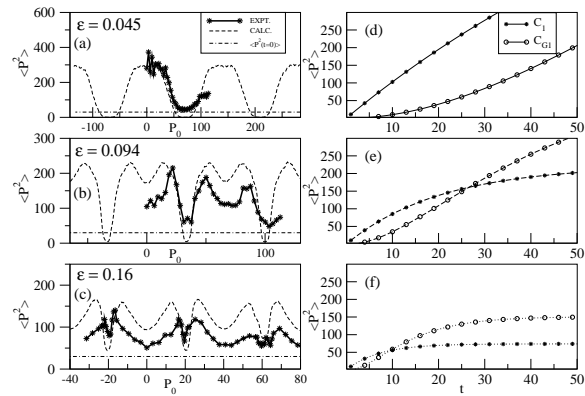}
\caption{Experimental realization of the $2\delta$-KP with a cloud
of cold cesium atoms, pulsed periodically by pairs of laser kicks.
The graphs show energy absorption of the ensemble for $K=3.3$,
$\epsilon=0.045,0.094,0.160$ and $\hbar=1$. Measurements are
taken after the quantum break time, $t^{*}\sim 30$, when the
energies have saturated and momentum diffusion has been terminated.
Varying $K_{\epsilon}$ means that the timescale $t_G$ on which the
global diffusion corrections become dominant varies in relation to
$t^{*}$. In Fig.\ref{Fig7}a $t^{*}<<t_G$ and the system is arrested
in a regime where $C_1$ dominates, corresponding to Fig.\ref{Fig5}a.
Atoms prepared in the trapping regions remain trapped and absorb no
energy. In Fig.\ref{Fig7}b $t^{*}\simeq t_G$ and the effects of the
global corrections $C_{Gn}$ become visible: inverted Poisson peaks
appear at trapping momenta and a partial sign reversal of the
overall cosine envelope is seen. In Fig.\ref{Fig7}c finally,
$t^{*}>t_G$ and the system follows classical energy trajectories
long enough for the global corrections to become dominant. Trapped
atoms start to absorb energy and diffusion in other parts of phase space
approaches a $(-\cos p_0 \epsilon)$ relationship. Figures d-f show
the behavior of $C_1$ (stars) and $C_{G1}$ (circles) with time for
the values of $K_{\epsilon}$ in a-c ($\tau=2-\epsilon$). We can
consider $t_G$ to be near the crossing point of the two correlations
shown. This clearly shifts to earlier times as $K_{\epsilon}$
increases.} \label{Fig7}
\end{figure}

Fig.\ref{Fig7} shows the energy absorption of a cloud of cold atoms
for $K=3.3$ and $\hbar=1$, measured in the $2\delta$-KP experiment
in \cite{PRL}. Values of $\epsilon$ vary from $0.045$ to $0.160$
such that $K_{\epsilon}=0.1485, 0.3102, 0.528$. The resemblance to
Fig.\ref{Fig5} is evident and is due to these different values of
$K_{\epsilon}$ which control the relative importance of diffusion
correlations. It should be noted that in experiments, $\epsilon$ and
$\tau$ are usually dimensionless scaled quantities, such that
$\epsilon=\Delta/T$, where $\Delta$ is the physical time between
kicks in a pair and $T$ is the physical period of the double-kick
system. The measurements are all taken well after the break time
$t^{*}\simeq D_0\sim 30$, when the energies have saturated and no
further evolution takes place. The experimental behavior then
depends on which diffusion corrections are dominant at the break
time. As $K_{\epsilon}$ decreases the time needed for the global
diffusion corrections to dominate over $C_1$ increases, hence in
Fig.\ref{Fig7}a the system is arrested at a time where $C_1$ is
still hugely dominant. In Figures \ref{Fig7}b and \ref{Fig7}c
however, $K_{\epsilon}$ is larger and the global correlations
$C_{Gn}$ become more important when the break time is reached. In
these cases the Poisson peaks and sign reversal for the cosine
envelope are clearly seen. Figures \ref{Fig7}d-\ref{Fig7}f show
$C_1$ and $C_{G1}$ for the parameters in a-c, and the crossing
point, indicating the point at which the global correlations become
dominant, clearly shifts from $t_G>t^{*}$ to $t_G<t^{*}$ at higher
$K_{\epsilon}$. Saturation values of correlations also decrease with
increasing time between the double kicks, which results in the limit
of the Standard Map where $\epsilon=\tau\simeq 2$ and there is no
momentum dependent diffusion at all. Note that at very high values
of $K$, and hence $K_{\epsilon}$, correlations become less
significant due to large Bessel arguments ($J_n(x)\to 0$ as
$x\to\infty$).

From the momentum distribution of the atomic ensemble one can also
observe the trapping regions and the effects of the global diffusion
corrections. After the quantum break time, the distribution of the
usual $\delta$-KP localizes exponentially in the momentum basis,
causing the characteristic triangular shape on a logarithmic plot.
For the $2\delta$-KP the basic exponential shape has a `staircase'
superposed onto it (see \cite{critstats,QKR2}).

\section{Asymptotic Diffusion: $t \to \infty$}

Classically, the momentum-dependent diffusion in the $2\delta$-KP does not
continue indefinitely but is only a transient effect at short and
intermediate times. At very long times in the asymptotic regime,
momentum-dependent correlations saturate and the diffusion rate is
the same for all starting conditions $p_0$. In this regime diffusion is controlled by
linear corrections independent of initial momentum, as for the
Standard Map, which decrease in magnitude with increasing separation
between kicks but remain constant in time.

Such corrections are obtained from Eq.(\ref{eq12}) in the case where the
total sum of $m_j$ coefficients is zero; hence $n=0$ and $\cos n p_0
\epsilon = 1$. It is found that all these corrections are of
opposite sign to the random walk $D_0=K^2$ and hence lower the
overall rate of energy absorption for all values of $K_{\epsilon}$.
This is due to the absence of terms for which $|m^{(2)}_j|=2$ or
$m^{(1)}_j\neq 0$. Although for momentum-independent correlations,
terms can also depend on the long timescale $\tau$ between kick
pairs, ie $m_j^{(1)}=\pm 1$, similar to the Standard Map case
($\tau\simeq T$), such terms are far less significant than the
dominant $\epsilon$-dependent corrections. Hence k-kick correlations
for even $k$, such as the important 2-kick correction ($\propto
J_2$), are effectively absent in the $2\delta$-KP. The general form
of correlations dependent on $\epsilon$ only is
\begin{eqnarray}
C_{G0}t=-K^2[\prod_m J^2_1(m K_{\epsilon})]
\sum^t_j\sum_{\alpha_m=0}^{f(j)} (\prod_m J_0^{2\alpha_m}(m
K_{\epsilon})) \nonumber \\
\label{eq21}
\end{eqnarray}
Note the difference in sign in comparison to $D_0$ and absence of
Bessel functions of order $p>1$. Note further that although we
denote these correlations by $Ct$, the actual diffusion rate
$D(t\to\infty)$ is non-zero and can be found by taking the infinite
time limit of \ref{eq21}. From the lowest order correlation for
which $m_j=\pm 1$, $m_k=\mp 1$ ($k<j$) we obtain a global family
similar to the case of the Standard Map, including the usual lowest
order 3-kick correlation
\begin{eqnarray}
C_{G0}t(\pm 1,\mp 1)=-K^2
J_1^2(K_{\epsilon})\sum_{j=2}^t\sum_{\alpha=0}^{j-2}
J_0^{2\alpha}(K_{\epsilon}) \label{eq22}
\end{eqnarray}
and hence, for $K_{\epsilon} \gg 1$,
\begin{eqnarray}
D \approx K^2 [1-\frac{J_1^2(K_{\epsilon})}{1-J_0^2(K_{\epsilon})}]
\label{eq23}
\end{eqnarray}
Note the similarity to the lowest order global family $C_k$
presented for the Standard Map earlier. The minor difference of a
$J_0^2$ factor stems from the fact that this family includes only
the case of odd $k$. Higher order corrections for the $2\delta$-KP
have the asymptotic form $D=-K^2\prod_m ((J_1^2(m
K_{\epsilon}))/(1-J_0^2(m K_{\epsilon})))^n$ for some $m,n$.

\begin{figure}[htb]
\includegraphics[width=3.3in]{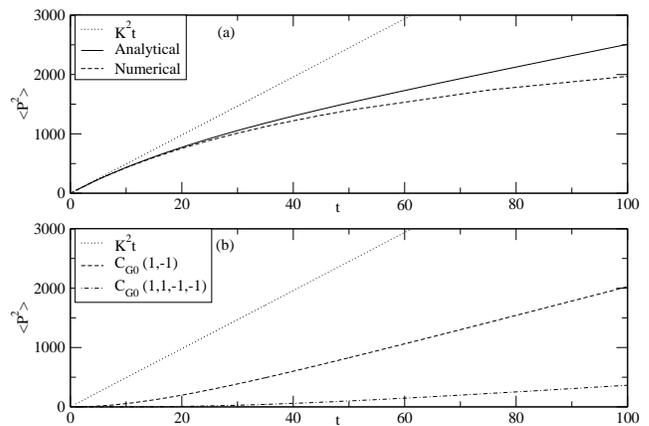}
\caption{Effect of momentum-independent corrections to the overall
rate of energy absorption by the $2\delta$-KP. Fig.\ref{Fig8}a shows
the significant decrease in energy absorption from the basic random
walk, $D_0=K^2$. The analytical curve is composed of the corrections
in Fig.\ref{Fig8}b and shows reasonable agreement with the numerical
data. Parameters are as in Fig.\ref{Fig5}. Absolute values of the
correlations are plotted in \ref{Fig8}b, so the analytical curve in
\ref{Fig8}a is $K^2t-C_{G0}t (\pm 1,\mp 1)-C_{G0}t (\pm 1,\pm 1,\mp 1,\mp 1)$.}
\label{Fig8}
\end{figure}

Fig.\ref{Fig8}a shows the effect of the two lowest order global
correlation families on the energy absorption of the kicked
particles for the same parameters as in Fig.\ref{Fig5}
($K_{\epsilon}=0.35$). The change from the $K^2$ random walk is seen
to be very significant. It should be noted that at early times
$C_{G0}t$ behaves non-linearly; however, this is only due to the
inclusion of entire global families of correlations between all
kicks. As time increases, more correlations are included in the
formula, however each individual k-kick correction still has a
linear time-dependence, as for all correlations independent of
momentum. At long times the energy increase becomes linear again,
since corrections become weaker with increasing $k$.

Fig.\ref{Fig8}b shows absolute values of the two corrections
included in the analytical curve in \ref{Fig8}a in comparison to
$D_0=K^2$. It is seen that the lowest-order term given in \ref{eq22}
makes up the vast majority of the diffusion correction. The second
term (given in the appendix) is derived from a sequence of $m_j=\pm
1, \pm 1, \mp 1, \mp 1$ and is hence $O(J_1^6)$. The agreement
between the analytical and numerical data in \ref{Fig8}a is
reasonable and can be improved by including higher-order terms.

\begin{figure}[tb]
\includegraphics[width=3.0in]{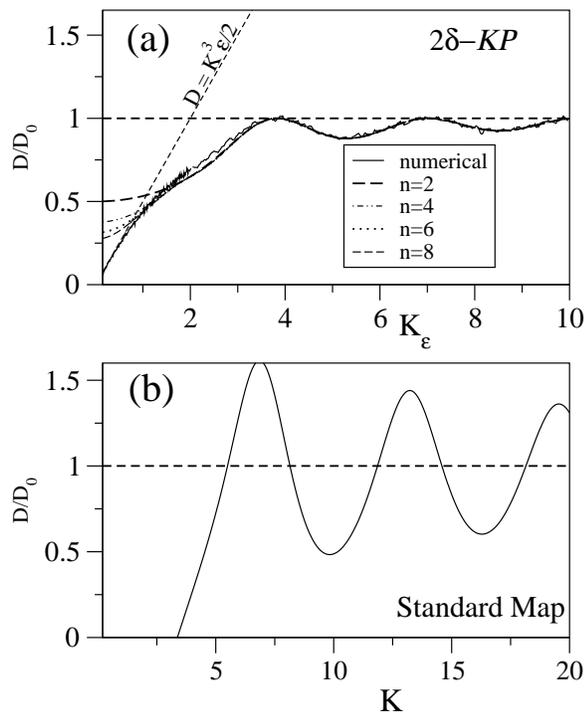}
\caption{ {\bf (a)} The diffusion rate for the $2\delta$-KP (shown as a function
of stochasticity parameter $K_\epsilon = K\epsilon$) never exceeds the 
`random walk' rate $D_0$; the uncorrelated rate,
(for time counted in kick-pairs), is $D_0=K^2$.
The figure shows that for large $K_\epsilon$, there is excellent agreement
between Eq.\ref{eq23} and numerics; in the regime of cold atom experiments,
$K_\epsilon \lesssim 1$, dominated by fractal phase-space structures,
agreement with the numerical result is improved
by including higher order correlations ($O(J_1^n)$) but remains non-quantitative.
Note that the numerical result $D \simeq K^3 \epsilon/2$ is represented 
by a straight (dashed) line since $D/D_0 = K_\epsilon/2$)\\
{\bf (b)} Illustrates the `textbook'  Standard Map diffusion rate
 $D \simeq K^2/2[1-2J_2(K)...]$ obtained by Rechester and White \cite{RechWhite}.
The rate oscillates about the `random walk'  value and the maxima
occur where classical accelerator modes are important.} \label{Fig9}
\end{figure}

Fig.\ref{Fig9} shows a comparison of the ratio $D/D_0$ between the
Standard Map and the $2\delta$-KP. Fig.\ref{Fig9}b shows the
well-known diffusion behavior for the case of the Standard Map or
$\delta$-KP where $D_0=K^2/2$ and $D$ is given by \ref{eq11} (this
curve was first presented in \cite{RechWhite} but without the
$C_4=K^2 J_2^2(K)$ term). $D/D_0$ vs $K$ is shown and the behavior
is oscillatory around the usual random walk with some regimes of
enhanced and some of hindered diffusion. At very large $K$ the
diffusion approaches the random walk value asymptotically. As was
mentioned in Section II, higher order global corrections do not
alter the shape of the curve in \ref{Fig9}b appreciably. For
$K\lesssim 5$ a linear relationship is found where $D\propto K^3$,
however as $K$ is decreased the system becomes increasingly regular
and the diffusion corrections  no longer describe
the behavior accurately.

Fig.\ref{Fig9}a shows $D/D_0$ vs $K_{\epsilon}$ for the
$2\delta$-KP, where $D_0=K^2$ (in unscaled momenta) and a series of
analytical curves are shown, including an increasing number of
global correction families, in comparison to the numerically
obtained diffusion behavior. The analytical curves are denoted by
the maximum order of $J_1(m K_{\epsilon})$ included. The behavior
for large $K_{\epsilon}$ is found to be different from the case of the
Standard Map (seen in \ref{Fig9}b):  for the $2\delta$-KP
  the diffusion is always `hindered'
compared to the random walk; ie $D<D_0$ for all $K_{\epsilon}$,
while for the Standard Map, there are maxima in the diffusion where
$D >D_0$. For the latter, the maxima correspond to values
of the stochasticity parameter $K \approx 2\pi M$ where $M$ is
an integer. These coincide with the appearance of small transporting
islands, (ie trajectories in these islands gain $\approx 2 \pi$
in momentum for each kick), the so-called ``accelerator modes''
 see eg \cite{Fishaccel}; the surrounding chaotic phase space generates
Levy flights. This regime of anomalous diffusion was observed
in the experiments in \cite{Raizen}.
We have found no evidence whatsoever of any accelerator modes
for $K \approx 2\pi M$; we cannot exclude a small contribution
from other higher-order accelerator modes, but the small maxima
of the $D/D_0$ in the $2\delta$-KP correspond to zeros of $J_1(K_{\epsilon})$
and the diffusion is bounded from above by the random walk rate.
Asymptotically, $D\to D_0$ as $K_{\epsilon}\to\infty$ as all correlations
decay away completely.

 For small
$K_{\epsilon}$ we again have a near-linear section where $D\propto
K^3\epsilon$ (or $D\propto K^3_{\epsilon}$ in scaled momenta
$p^{\epsilon}$) as mentioned in Section II. 
The agreement
between analytical and numerical results is excellent for
$K_{\epsilon}\gtrsim 2$, where even the lowest order ($O(J_1^2)$)
correlations are sufficient to reproduce the diffusion curve. As
$K\to 0$ agreement can be improved by including successively higher order
corrections.

\section{CONCLUSIONS}

We have analyzed the chaotic classical diffusion of the
double-$\delta$-KP. Although  a
straightforward extension of the Standard Map, this system
 exhibits a rich variety of new features. 

One motivation of the present work was to
further understand the behavior observed in experiments \cite{PRL} 
and to present a more detailed derivation of the formulae in \cite{PRL}.
The experiment probed regimes where the classical diffusion is highly non-linear
in time and depends on initial conditions, but which were nevertheless
sufficiently generic to be analysed in terms of corrections to a
diffusive process. This is in contrast to the Standard Map, where
long-range correlations are a feature of the near-integrable regime
where most trajectories are stable so analysis based on diffusive
processes is not useful.
Although the non-linear regime of the $2\delta$-KP is, classically, transient, it is
sufficiently long-lived ($\sim 100-200$ kicks) so that it is the {\em only}
regime sampled by the quantum dynamics of the experiment in \cite{PRL}.

Quantum studies of the $2\delta$-KP in the regime $0.1 \lesssim K_\epsilon \lesssim 1$
were originally undertaken because they coincide roughly with the experimental
values. However, the theoretical studies then revealed novel quantum behaviour
in the energy level-statistics and the fractional scalings $L \sim \hbar^{-0.75}$
of the quantum localization lengths \cite{QKR2,critstats} . It was
 suggested that the {\em global}  quantum properties of the $2\delta$-KP 
are, in this regime,
intimately connected with the {\em local} scalings around the neighborhood of the
Golden ratio cantori, as transport from cell to cell is limited by the
former. The $K^3$ dependence of the diffusion rate found here lends further
credence to this suggestion, as a cubic dependence is a feature of diffusion near 
Golden-ratio cantori \cite{LichLieb}. Unfortunately, as for the Standard map, 
the cubic behaviour cannot be derived analytically. At best, it was seen that adding successively
higher orders of diffusive corrections makes the analysis tend towards the numerics.

One of the main new results here is contained in Fig.\ref{Fig9}, and Eq.\ref{eq23}:
in the regime $K_\epsilon \gg 1$, where effects
of cantori and trapping  are negligible. Here, the diffusion of the $2\delta$-KP follows 
a simple analytical expression but
still differs strikingly from the Standard Map, in that it never exceeds the
random walk rate (though it can equal it). The $2\delta$-KP has no two-kick correction
(the $J_2(K)$ term  which in the Standard Map roughly `tracks' the accelerator
modes). We have found no numerical evidence for accelerator mode behavior. Hence, although there
is much evidence for non-linear diffusion (of finite though prolonged duration)
 in the $2\delta$-KP, the major source of anomalous diffusion seen in the chaotic 
Standard Map is in fact absent; in this sense, $2\delta$-KP  diffusion is closer to 
the random walk (as shown in Fig.9) once trapping by cantori becomes unimportant.

Finally, it is worth noting the potential applications of the
$2\delta$-KP, in particular as a velocity-selective atom filter in,
for example, devices like an atom chip. Narrow trapping
regions could be used to select atoms with $p_0\simeq \pi/\epsilon$,
while others would accelerate through the system nearly unperturbed.
This could also be used to create very pure Bose-Einstein
condensates, if located in a trapping region. A much stronger
momentum-dependent effect is seen in the double-kick system than in
the previously studied perturbed-period system \cite{PPRtheory}.
Other applications in atomic manipulation may also be possible given
further investigations.
\\
\\
 M Stocklin acknowledges a Graduate Research Scholarship; we acknowledge
helpful discussions with Shmuel Fishman and Charles Creffield. 
\\

\begin{widetext}
\appendix*
\section{GLOBAL CORRELATION FAMILIES}

Exact forms of two of the momentum-dependent global correlation
families presented in this paper are given here as examples, as well
as one further example of a higher order $C_{G0}$ correlation term.

Note that the $m_j^{(2)}$ patterns given here and also shown in
Table 1 correspond to the sum of correlations with such non-zero
coefficients, including all possible permutations. The sum of the
coefficients defines the cosine order $n$ and permutations for which
the partial sum, when coefficients are added from highest to lowest
$j$, is zero, lead to zero-valued correlations.

$m_j=\pm 1, \pm 1, \mp 1$ (Term Ib in Table 1)

\begin{eqnarray}
C_{G1}^{(1)}t=-K^2 \cos p_0 \epsilon J_1^3(K_{\epsilon})
J_1^2(2K_{\epsilon}) [3\frac{J_0(K_{\epsilon})-J_2(K_{\epsilon})}
{J_1(K_{\epsilon})}+2\frac{J_0(2K_{\epsilon})-J_2(2K_{\epsilon})}{J_1(2K_{\epsilon})}]\nonumber \\
\times \sum_{j=3}^t \sum_{\alpha=0}^{j-3}
J_0^{2\alpha}(K_{\epsilon}) J_0^{2(j-3-\alpha)}(2K_{\epsilon})
\label{eq24}
\end{eqnarray}

$m_j=\pm 1, \pm 1$ (Term IIa in Table 1)

\begin{eqnarray}
C_{G2}^{(0)}t=2K^2 \cos 2p_0 \epsilon J_1(2K_{\epsilon})
J_1^2(K_{\epsilon}) \sum_{j=2}^t
\sum_{\alpha=0}^{j-2}(2\alpha\frac{J_1(K_{\epsilon})}
{J_0(K_{\epsilon})}+2(j-2-\alpha)\frac{J_1(2K_{\epsilon})}{J_0(2K_{\epsilon})})
J_0^{2\alpha}(K_{\epsilon}) J_0^{2(j-2-\alpha)}(2K_{\epsilon})
\label{eq25}
\end{eqnarray}

$m_j=\pm 1, \pm 1, \mp 1, \mp 1$ (second higher-order correlation
term in Fig.\ref{Fig8}b and included in Fig.\ref{Fig9}a)

\begin{eqnarray}
C_{G0}t(\pm 1,\pm 1,\mp 1,\mp 1)=-K^2[J_1^4(K_{\epsilon})J_1^2(2K_{\epsilon})]
\sum_{j=4}^t\sum_{\alpha_1=0}^{j-4}\sum_{\alpha_2=0}^{j-4-\alpha_1}
\sum_{\alpha_3=0}^{j-4-(\alpha_1+\alpha_2)}
J_0^{2\alpha_1}(K_{\epsilon}) J_0^{2\alpha_2}(2K_{\epsilon}) J_0^{2\alpha_3}(K_{\epsilon})
\label{eq26}
\end{eqnarray}

Other Poisson and higher-order global correlation terms have similar
forms and are derived in the same way as the above from \ref{eq12}.
Note that the time behavior ($C_{Gn}^{(1)}$ terms increase as $\sim
t^n$, while $C_{Gn}^{(0)}$ terms increase as $\sim t^{n+1}$) for
small $K_{\epsilon}$ and $t$, can be shown for all terms by
evaluating the sequence of geometric sums and expanding powers
binomially to the appropriate order. Saturation values and
asymptotic behavior are also straightforward to evaluate by taking
the limit as $t\to\infty$.
\end{widetext}




\begin{thebibliography}{99}
\bibitem{map} The $\delta$-KP (\ref{eq10}) does not include a $K\sin x_t$ term,
so in theory we cannot have a correlation with the kick at time $t$, since
$m_t=\pm 1$ produces a term which is zero. For the $2\delta$-KP
(\ref{eq12}) we have changed the $S$ notation slightly so that a
$K\sin x_t^{(2)}$ term is included at the expense of $K\sin x_0$,
where we can choose to set $x_0=0$. Either notation does not change
the underlying physics and boundary effects at the start and finish
of the system evolution can be neglected.
\end{thebibliography}
\end{document}